\documentclass[pra,twocolumn,showpacs]{revtex4}
\usepackage{graphicx}
\begin{document}
\title{Decoherence-free quantum dynamics in circuit QED system}
\author{Ping Liao$^{1}$, Qin-Qin Wu$^{1}$, Jie-Qiao Liao$^{1,2}$, and  Le-Man Kuang$^{1}$\footnote{Author to whom any correspondence should be
addressed. }\footnote{ Email: lmkuang@hunnu.edu.cn}}
\address{$^{1}$Key
Laboratory of Low-Dimensional Quantum Structures and Quantum
Control of Ministry of Education, and Department of Physics, Hunan
Normal University, Changsha 410081, People's Republic of China\\
$^{2}$Institute of Theoretical Physics, Chinese Academy of
Sciences, Beijing, 100080, China}

\begin{abstract}
We study decoherence in a circuit QED system consisting of  a
charge qubit and two superconducting transmission line resonators
(TLRs). We show that in the dispersive regime of the circuit QED
system one TLR can be used as an auxiliary subsystem to realize
decoherence-free quantum dynamics of the bipartite target system
consisting of the charge qubit and the other TLR conditioned on
the auxiliary TLR initially being a proper number state. Our study
gives new insight into control and manipulation of decoherence in
quantum systems.
\end{abstract}

\pacs{03.65.Yz, 03.67.Mn, 03.67.Lx} \maketitle

\section{Introduction}
Decoherence remains a major obstacle to experimental realizations
of quantum computation and communication \cite{1,2}. As is well
known, no system can be completely isolated from its environment.
Interactions between the system and the environment create
decoherence. There are some interesting methods to bypass
decoherence in quantum information processing. One of them is to
encode quantum information into decoherence-free subspaces and
subsystems (DFSs) \cite{3,4,5,6,7,8,9}. Under certain conditions,
a subspace of a physical system is decoupled from its environment
such that the dynamics within this subspace is purely unitary. The
DFS is a set of all states which is not Experimental realizations
of DFSs have been achieved in photon systems \cite{10,11,12,13},
nuclear spin systems \cite{14,15,16}, and trapped-ion systems
\cite{17,18}.

Advances in circuit QED [19-42] opened new  prospects in
nonclassical state generation and quantum information processing
in the microwave regime. In the circuit QED, superconducting
circuits are made to act like artificial atoms and a
one-dimensional superconducting transmission line resonator (TLR)
forms a microwave cavity. Unlike natural atoms, the properties of
artificial atoms made from circuits can be designed to taste, and
even manipulated in-situ. Because the qubit contains many atoms,
the effective dipole moment can be much larger than an ordinary
alkali atom and a Rydberg atom. This allows circuits to couple
much more strongly to the cavity. This large coherent coupling
allows circuits to achieve strong coupling even in the presence of
the larger decoherence present in the solid state environment,
then one can observe the quantum interactions of matter with
single photons. Hence, circuit QED can explore new regimes of
cavity QED. Recently, the circuit QED systems have successfully
demonstrated strong coupling between a single microwave photon and
a qubit \cite{20}, the implementation of a single microwave-photon
source in all solid-state system \cite{21}, as well as single
artificial-atom lasing \cite{22} and interaction between two
artificial atoms \cite{23,24}. More recently, the Lamb shift,
two-photon Jaynes-Cummings model and controlled symmetry breaking
\cite{25} have also been observed experimentally in circuit-QED
systems \cite{26,27}. These give rise to strong experimental
supports for on-chip quantum optics and quantum information
processing.

In this paper, we are concerned with decoherence in a circuit QED
system which includes one SQUID-type charge qubit acting as an
artificial atom and two superconducting TLRs. We show that in the
dispersive regime of the circuit QED system one TLR can be used as
an auxiliary subsystem to realize decoherence-free quantum
dynamics of the bipartite target system consisting of the charge
qubit and the other TLR.  The paper is organized as follows. In
Sec. II, we propose the physical model under our consideration and
present its analytical solution in the dispersive regime. In Sec.
II, we investigate DFS of the circuit-QED system. We show how to
realize  decoherence-free quantum dynamics of the bipartite target
system. We shall conclude the paper with discussions and remarks
in the last section.

\begin{figure} [htp]
\center
\includegraphics[width=8.3cm,height=4.35cm] {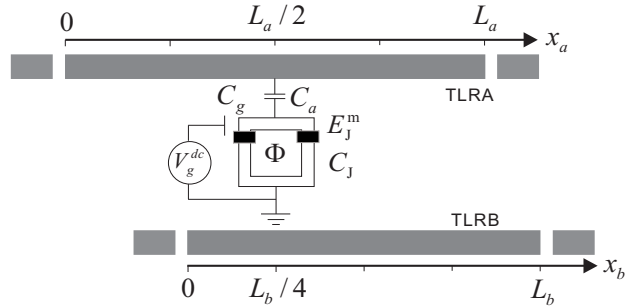}
\caption{Schematic setup for the proposed circuit-QED system. The
SQUID-based charge qubit is coupled with two TLRs, TLRA and TLRB,
of lengths $L_{a}$ and $L_{b}$, respectively. The SQUID is placed
at the position of the antinode of the quantized voltage of TLRA
(i.e., $L_{a}/2$) and the antinode of the quantized current of
TLRB (i.e., $L_{b}/4$), respectively. }
\end{figure}

\section{Physical model}
Let us illustrate our idea firstly. As shown in Fig.1, we consider
a circuit-QED system in which an SQUID-type charge qubit is
coupled to two transmission line resonators, TLRA and TLRB, of
lengths $L_{a}$ and $L_{b}$, respectively. The qubit is placed at
the position of the antinode of the quantized voltage of TLRA
(i.e., $x_a=L_{a}/2$) and the antinode of the quantized current of
TLRB (i.e., $x_b=L_{b}/4$), respectively. It can be controlled by
the gate voltage, which contains the dc part $V_{g}^{dc}$ and the
quantum part $V_{a}$ generated by the TLRA, and the biasing flux
$\Phi$, which contains the classical part $\Phi_{e}$ and the
quantized part $\Phi_{b}$ generated by the TLRB threading the
SQUID.

In terms of the annihilation operator $a (b)$ and creation
operator $a^{\dag} (b^{\dag})$ of TLRA (TLRB), the Hamiltonian for
this system reads as \cite{19,20}
\begin{eqnarray}
\label{e1}
H=\hbar\omega_{a}a^{\dag}a+\hbar\omega_{b}b^{\dag}b+2E_{C}(2n_{g}-1)\sigma_{z}
-E_{J}\sigma_{x},
\end{eqnarray}
where $\omega_{a}$ and $\omega_{b}$ are the microwave frequencies
of TLRA and TLRB, respectively. The last two terms represent the
Hamiltonian of the charge qubit. Here
$\sigma_{z}=|1\rangle\langle1|-|0\rangle\langle0|$ and
$\sigma_{x}=|1\rangle\langle0|+|0\rangle\langle1|$ with $0(1)$ the
number of Cooper pairs on the superconducting island.
$E_{C}=e^{2}/2C_{\Sigma}$ is the charging energy with $C_{\Sigma}$
being the total box capacitance. $n_{g}$ is the gate charge number
and $E_{J}$ is the Josephson coupling energy given by
\begin{eqnarray}
\label{e2}
n_{g}=\frac{C_{g}V_{g}^{dc}+C_{a}V_{a}}{2e},\;\;\;\;\;\;\;E_{J}=E_{J}^{m}\cos\left(\pi\frac{\Phi}{\Phi_{0}}\right),
\end{eqnarray}
where $C_{g}$ and $C_{a}$ represent the gate capacitance and the
coupling capacitance between TLRA and the charge qubit.
$V_{g}^{dc}$ and $V_{a}$ are the dc gate voltage and the quantum
gate voltage generated by TLRA, respectively. $E_{J}^{m}$ is the
maximum Josephson coupling energy and $\Phi_{0}$ is the flux
quanta. The total magnetic flux $\Phi$ threading the dc-SQUID is a
sum of two parts $\Phi=\Phi_{b}+\Phi_{e}$ with $\Phi_{e}$ being
the external classical magnetic flux and $\Phi_{b}$ the quantized
magnetic flux generated by the quantized current in TLRB.

The quantum gate voltage  and the quantized magnetic flux
associated with  TLRA and TLRB can be expressed in terms of the
annihilation and creation operators of the microwave fields in
TLRA and TLRB as
\begin{eqnarray}
\label{e3}
V_{a}&=&-\sqrt{\frac{\hbar\omega_{a}}{L_{a}c}}(a+a^{\dag}),\nonumber\\
\Phi_{b}&=&\frac{\mu_{0}S}{2\pi
d}\sqrt{\frac{\hbar\omega_{b}}{L_{b}l}}(b+b^{\dag}),
\end{eqnarray}
where $c$ and $l$ are the capacitance and inductance per unit
length for TLRA and TLRB, respectively, $S$ is the area of the
loop of the SQUID, and $d$ the distance between TLRB and the SQUID
and $\mu_{0}$ the vacuum permeability.

Substituting Eqs. (\ref{e2}) and (\ref{e3}) into Eq. (\ref{e1}) we
get
\begin{eqnarray}
\label{e4}
H&=&\hbar\omega_{a}a^{\dag}a+\hbar\omega_{b}b^{\dag}b+2E_{C}(2n_{g}^{dc}-1)\sigma_{z}\\
&&-\hbar
g_{a}(a+a^{\dag})\sigma_{z}-E_{J}^{m}\cos[\phi_{e}+\phi_{b}(b+b^{\dag})]\sigma_{x}\nonumber,
\end{eqnarray}
where $n_{g}^{dc}=C_{g}V_{g}^{dc}/(2e)$ and we have introduced the
coupling constant $g_{a}$,  two parameters $\phi_{b}$ and
$\phi_{e}$. They are defined by
\begin{eqnarray}
\label{e5}
g_{a}&=&2E_{C}C_{a}\sqrt{\hbar\omega_{a}/(L_{a}c)}/(\hbar e),
\nonumber\\
\phi_{b}&=&\mu_{0}S\sqrt{\hbar\omega_{b}/(L_{b}l)}/(2d\Phi_{0}),\nonumber\\
\phi_{e}&=&\pi\Phi_{e}/\Phi_{0}.
\end{eqnarray}

For the simplicity, we choose the classical biasing magnetic flux
$\Phi_{e}=0$ and  work at the charge degeneracy point
$n_{g}^{dc}=1/2$. After making a rotation of $-\pi/2$ around the
$y$ axis, we get the following effective Hamiltonian
\begin{eqnarray}
\label{e6} H^{'}&=&\hbar\omega_{a}a^{\dag}a+\hbar\omega_{b}b^{\dag}b-\hbar g_{a}(a+a^{\dag})\sigma_{x}\\
&&+E_{J}^{m}\cos[\phi_{b}(b+b^{\dag})]\sigma_{z}\nonumber,
\end{eqnarray}
which indicates that under the condition $\phi_{b}\ll 1$, we can
obtain the following approximation Hamiltonian
\begin{eqnarray}
\label{e7}
H^{''}&=&\hbar\omega_{a}a^{\dag}a+\hbar\omega_{b}b^{\dag}b+E_{J}^{m}[1-\phi_{b}^{2}(1+2b^{\dag}b)/2]\sigma_{z}\nonumber\\
&&-\frac{E_{J}^{m}\phi_{b}^{2}}{2}(b^{2}+b^{\dag2})\sigma_{z}-\hbar
g_{a}(a+a^{\dag})\sigma_{x}.
\end{eqnarray}

In order to further simplify the above Hamiltonian, we change
Hamiltonian (\ref{e7}) to the interaction picture with respect to
the free Hamiltonian of TLRB. After discarding rapidly oscillating
terms, the resulting Hamiltonian can be expressed as
\begin{equation}
\label{e8}
H^{''}_I=\hbar\omega_{a}a^{\dag}a+\frac{\hbar\omega_{q}}{2}\sigma_{z}-\hbar
g_{a}(a+a^{\dag})\sigma_{x},
\end{equation}
where the effective energy separation of the qubit is dependent of
the number operator of TLRB $n_{b}=b^{\dag}b$, and it is given by
the following expression
\begin{equation}
\label{e9}
\omega_{q}=2E_{J}^{m}[1-\phi_{b}^{2}(1+2n_{b})/2]/\hbar,
\end{equation}

We consider the case of
$\omega_{q}+\omega_{a}\gg\omega_{q}-\omega_{a}, g_{a}$. Then under
the rotating-wave approximation  Hamiltonian (\ref{e7}) becomes
\begin{equation}
\label{e10}
H^{''}_I=\hbar\omega_{a}a^{\dag}a+\frac{\hbar\omega_{q}}{2}\sigma_{z}-\hbar
g_{a}(a\sigma_{+}+\sigma_{-}a^{\dag}),
\end{equation}
which is a generalized Jaynes-Cummings model which describes the
interaction between TLRA and the charge qubit with the effective
energy separation depending on the number operator of TLRB. The
quantum electric circuit of Fig.1 is therefore mapped to the
problem of a two-level artificial atom inside a cavity.

We study the dispersive regime of the circuit QED, where the
cavity and the qubit are out of resonance, and the qubit-cavity
detuning is larger than the coupling strength, i.e.,
$\Delta=\omega_{q}-\omega_{a}\gg g_{a}$. In the dispersive regime,
$\omega_{q}/\omega_{a}<1$ from  Hamiltonian (\ref{e9}) we can
obtain the following effective Hamiltonian
\begin{eqnarray}
\label{e11}
H^{'''}&=&\hbar\omega_{a}a^{\dag}a+\frac{\hbar\omega_{q}}{2}\sigma_{z}
-\hbar\frac{g_{a}^{2}}{\omega_{a}}\left(1+\frac{\omega_{q}}{\omega_{a}}\right)\nonumber\\
&&\times[\sigma_{z}a^{\dag}a+(\sigma_{z}+1)/2],
\end{eqnarray}
which can be expressed as the following form in the interaction
picture with respect to the first two terms of the Hamiltonian
\begin{eqnarray}
\label{e12} H_S&=&H_0|0\rangle\langle 0| + H_1|1\rangle\langle 1|,
\end{eqnarray}
where $H_0$ and  $H_1$ are defined by
\begin{eqnarray}
\label{e13} H_{0}&=&\hbar\omega'_{a}a^{\dag}a - \hbar\chi
a^{\dag}a b^{\dag}b, \nonumber\\
H_{1}&=&-\hbar\omega'_{a}(a^{\dag}a+1)+ \hbar\chi b^{\dag}b +
\hbar\chi a^{\dag}a b^{\dag}b,
\end{eqnarray}
where we have introduced the effective frequency for the TLA and
the cross-Kerr coupling strength between TLA and TLB defined by
\begin{eqnarray} \label{e14}
\omega'_{a}&=&\frac{g_a^2}{\omega_{a}}+\frac{2g_a^2E^m_J}{\hbar\omega_{a}^2}-\frac{\chi}{2},
\nonumber\\
\chi&=&\frac{2g_{a}^{2}\phi_{b}^{2}E_{J}^{m}}{\hbar\omega^2_{a}},
\end{eqnarray}
which indicate that by a careful choice of the parameters, it is
possible to obtain considerable cross-Kerr nonlinearity. According
to recent experimental data in Ref. \cite{30}, $E_{C}/\hbar=5$GHz,
$E_{J}^{m}/\hbar=8$GHz, $\omega_{a}=2\pi\times6$GHz, $L_{a}=25$mm,
$C_{a}=c=0.13$fF, and $\phi_{b}=0.1$, we find the resulting
cross-Kerr coupling strength to be $\chi=360$MHz. In circuit QED,
the lifetime of the charge qubit and the transmission line cavity
[30] are about 2$\mu$s and 160ns, respectively. In the lifetime of
the transmission line resonator, $\tau=160$ns, we can reach a
cross-phase shift $\phi=\chi\tau\simeq 9.17\times 2\pi$. This
means that in the lifetime of the involved subsystems, we can
obtain a large cross-phase shift between two microwave fields in
the two transmission line resonators.

Obviously, Hamiltonian (\ref{e13}) is diagonal in the ($\sigma_z,
a^{\dag}a, b^{\dag}b$) representation with the following basis
\begin{eqnarray} \label{e15}
|m, n, i\rangle=\frac{1}{\sqrt{m!n!}}a^{\dag}b^{\dag}|m,
n\rangle\otimes|i\rangle,
\end{eqnarray}
where $n$ and $m$ take non-negative integers,  $i=0,1$, and
$|i\rangle$ are eigenstates of the qubit operator $\sigma_z$ with
$\sigma_z|0\rangle=-|0\rangle$ and $\sigma_z|1\rangle=|1\rangle$.
The states defined by Eq. (\ref{e15}) are eigenstates of the
Hamiltonian (\ref{e13}), i.e., $H_S|m, n, i\rangle=E_{nmi}|m, n,
i\rangle$ with the eigenvalues
\begin{eqnarray} \label{e16}
E_{mn0}&=&\hbar(\omega'_{a}-\chi n)m, \nonumber\\
E_{mn1}&=&-\hbar(\omega'_{a}-\chi n)(m+1).
\end{eqnarray}

\section{Decoherence-free subspace}
We now consider the decoherence dynamics of the circuit QED system
under our consideration. We use a reservoir consisting of an
infinite set of harmonic oscillators to model environment of the
circuit QED system and assume the total Hamiltonian  \cite{43,44}
to be
\begin{eqnarray}
\label{e17}
 \hat{H}_T&=&\hat{H}_S +
\sum_k\omega_k\hat{b}^{\dagger}_k\hat{b}_k +
\hat{H}_S\sum_kc_k(\hat{b}^{\dagger}_k+\hat{b}_k)\nonumber \\
&&+\hat{H}_S^2\sum_k\frac{c_k^2}{\omega_k^2},
\end{eqnarray}
where the second term is the Hamiltonian of the reservoir, the
third one represents the interaction between the system and the
reservoir with a coupling constant $c_k$, and the last one is a
renormalization term \cite{45}. Obviously, the interaction term in
Eq.(\ref{e17}) commutes with the Hamiltonian of the system, this
means that there is no energy exchange between the system and its
environment,  so that the decoherence  described by Hamiltonian
(\ref{e17}) is the phase decoherence. A previous work \cite{43}
has shown that a nonlinear extension of the model Hamiltonian
(\ref{e17}) can well describe the phase decoherence in trapped-ion
systems.

The Hamiltonian (\ref{e17}) can be exactly solved by making use of
the following unitary transformation
\begin{equation}
\label{e18}
 \hat{U}=\exp\left
[\hat{H}_S\sum_k\frac{c_k}{\omega_k}(\hat{b}^+_k-\hat{b}_k)\right
].
\end{equation}

Corresponding to the Hamiltonian (\ref{e17}), the total density
operator of the system plus reservoir can be expressed as
\begin{eqnarray}
\label{e19}
\hat{\rho}_T(t)&=&e^{-i\hat{H}_St}\hat{U}^{-1}e^{-it\sum_k\omega_k\hat{b}^+_k\hat{b}_k}\hat{U}\nonumber
\\
&&\times
\hat{\rho}_T(0)\hat{U}^{-1}e^{it\sum_k\omega_k\hat{b}^+_k\hat{b}_k}\hat{U}e^{i\hat{H}_St},
\end{eqnarray}
In the derivation of the above solution, we have used
$\hat{\rho}_T(t)=\hat{U}^{-1}\hat{\rho}'_T(t)\hat{U}$, where
$\hat{\rho}'_T=e^{-i\hat{H}'_Tt}\hat{\rho}'_T(0)e^{i\hat{H}'_Tt}$
with $\hat{H}'_T=\hat{U}\hat{H}_T\hat{U}^{-1}$ and
$\hat{\rho}'_T(0)=\hat{U}\hat{\rho}_T(0)\hat{U}^{-1}$, where
$\hat{\rho}_T(0)$ the initial total density operator.

We assume that the system and  reservoir are initially in thermal
equilibrium and  uncorrelated, so that
$\hat{\rho}_T(0)=\hat{\rho}(0)\otimes\hat{\rho}_R$, where
$\hat{\rho}(0)$ is the initial  density operator of the system,
and $\hat{\rho}_R$ the density operator of the reservoir, which
can be written as $\hat{\rho}_R=\prod_k\hat{\rho}_k(0)$ with
$\hat{\rho}_k(0)$ is the density  operator of the $k$-th harmonic
oscillator in thermal equilibrium. After taking the  trace over
the reservoir, from  Eq.(\ref{e18}) we can get the reduced density
operator of the system, denoted by
$\hat{\rho}(t)=tr_R\hat{\rho}_T(t)$, its matrix elements are
explicitly written as
\begin{eqnarray}
\label{e20}
\rho_{(m'n'i')(mni)}(t)=\rho_{(m'n'i')(mni)}(0)R_{(m'n'i')(mni)}(t)
\nonumber \\
\times\exp\left \{-i[E_{m'n'i'}-E_{mni}]t\right \},
\end{eqnarray}
where $R_{(m'n'i')(mni)}(t)$ is a reservoir-dependent quantity
given by
\begin{eqnarray}
\label{e21}
 R_{(m'n'i')(mni)}(t)&=&\prod_k
Tr_R\left\{\hat{V}^{-1}(E_{m'n'i'})e^{-it\omega_k\hat{b}^+_k\hat{b}_k}\right.
\nonumber
\\
& &\left. \times \hat{V}_k(E_{m'n'i'})
\hat{\rho}_k(0)\hat{V}^{-1}(E_{mni})\right. \nonumber
\\
& &\left. \times
e^{it\omega_k\hat{b}^+_k\hat{b}_k}\hat{V}_k(E_{mni})\right \},
\end{eqnarray}
with $\hat{V}_k(x)=\exp[xc_k(\hat{b}^+_k-\hat{b}_k)/\omega_k]$.

After somewhat lengthy but straightforward calculation, we find
that $R_{(m'n'i')(mni)}(t)$ can be factorized as
\begin{eqnarray}
\label{e22}
 R_{(m'n'i')(mni)}(t)&=& \exp\left [
-i\delta\phi_{(m'n'i')(mni)}(t)\right ]\nonumber
\\
&&\times \exp\left [-\Gamma_{(m'n'i')(mni)}(t)\right ].
\end{eqnarray}
Here the phase shift and the damping factor are defined by
\begin{eqnarray}
\label{e23}
 \hspace{-1cm}
\delta\phi_{(m'n'i')(mni)}(t)&=&\left(E^2_{m'n'i'}-E^2_{mni}\right) Q_1(t),\\
\label{e24} \Gamma_{(m'n'i')(mni)}(t)&=&(E_{m'n'i'}-E_{mni})^2
Q_2(t),
\end{eqnarray}
where the two reservoir-dependent functions are given by
\begin{eqnarray}
\label{e25} Q_1(t)&=&\int^{\infty}_{0} d\omega
J(\omega)\frac{c^2(\omega)}{\omega^2}\sin(\omega t),\\
\label{e26} Q_2(t)&=&2\int^{\infty}_{0} d\omega J(\omega)
\frac{c^2(\omega)}{\omega^2}\sin^2\left(\frac{\omega
t}{2}\right)\coth\left(\frac{\beta\omega}{2}\right).
\end{eqnarray}
where $J(\omega)$ is the spectral density of the reservoir, and
$\beta=1/k_BT$ with $k_B$ and $T$ being the Boltzmann constant and
temperature, respectively.

Eqs.(\ref{e20}) and (\ref{e22}) indicate that the interaction
between the system and its environment induces a phase shift
$\delta\phi_{(m'n'i')(mni)}(t)$ and a decaying factor
$\Gamma_{(m'n'i')(mni)}(t)$ in the reduced density operator of the
system.  All necessary information about the effects of the
environment is contained in the spectral density of the reservoir
\cite{45,46}.

Eqs.(\ref{e23}) and (\ref{e24}) we can see that the phase shift
and the decaying factor induced by the environment are determined
by the energy differences of the system under our consideration
$E_{m'n'i'}-E_{mni}$ and the spectral density of the environment
$J(\omega)$. This implies that decoherence  can be controlled and
manipulated by properly choosing the energy differences of the
system. In fact, From Eq. (\ref{e16}) we can see that there exist
three types of energy differences given by
\begin{eqnarray}
\label{e32}
 E_{m'n'1}-E_{mn0}&=&-\hbar\omega'_{a}(n+n'+1)+\hbar\chi m'\nonumber\\
 & &+\hbar\chi (n'm'+nm),\\
 \label{e33}
 E_{m'n'0}-E_{mn0}&=&\hbar\omega'_{a}(n'-n)-\hbar\chi (n'm'-nm),\\
 \label{e34}
 E_{m'n'1}-E_{mn1}&=&-\hbar\omega'_{a}(n'-n)+\hbar\chi(m'-m)\nonumber\\
 & &+\hbar\chi (n'm'-nm),
\end{eqnarray}
which indicates that when $m=m'=\omega'_{a}/\chi\equiv m_0$,  we
have $ E_{m_0 n' 1}-E_{m_0 n
0}=E_{m_0n'0}-E_{m_0n0}=E_{m_0n'1}-E_{m_0n1}=0$. Under these
conditions from Eq. (24) we can find that the damping factor
vanishes, i.e., $ \Gamma_{(m_0n'i')(m_0ni)}(t)=0$. This means that
the subspace $\{|m_0, n, 1\rangle, |m_0, n', 0\rangle\}$ with $n$
and $n'$ being arbitrary non-negative integers is a
decoherence-free subspace of the triple-partite circuit QED system
under our consideration. In other words, for our present
triple-partite system consisting one qubit and two TLRs if TLRA is
initially prepared the number state $|m_0\rangle$, then quantum
dynamics of the bipartite subsystem consisting of the charge qubit
and TLRB will be decoherence-free. This means that an arbitrary
quantum state of the bipartite system would be a decoherence-free
state conditioned on the auxiliary subsystem TLRA initially being
the number state $|m_0\rangle$. In this sense, TLRA acts as an
auxiliary subsystem which is used to control decoherence of the
bipartite system consisting of the charge qubit and TLRB. Hence,
we realize decoherence-free quantum dynamics of the charge qubit
and TLRB.

\section{Concluding remarks}
In conclusion, we have studied decoherence in the circuit QED
system consisting of  a charge qubit and two superconducting TLRs.
Actully, we have proposed a scheme to realize decoherence-free
quantum dynamics of the bipartite consisting of the charge qubit
and one superconducting TLR by using another superconducting TLR
as auxiliary subsystem. In this scheme one TLR and the charge
qubit constitute the controlled target system while the other TLR
is the auxiliary subsystem which acts as a tool to control the
target system.  The whole of them forms a triple-partite
circuit-QED system. It has been found that in the dispersive
regime of the circuit QED system, decoherence-free quantum
dynamics of the bipartite target system can be realized when the
auxiliary TLR subsystem is initially prepared in proper number
states. This implies that by controlling and manipulating the
auxiliary subsystem, one can protect quantum system against
decoherence. This provides fundamental insight into the control of
decoherence in circuit QED systems. It is believed that our
present scheme opens an new way to engineer decoherence in quantum
systems.

\acknowledgments This work was supported by the National
Fundamental Research Program Grant No.  2007CB925204, the National
Natural Science Foundation under Grant Nos. 10775048 and 10325523,
and the Education Committee of Hunan Province under Grant No.
08W012.


\end{document}